\documentclass[11pt]{article}
\usepackage{latexsym,amsmath,amssymb,theorem,epsfig}

\usepackage{amsfonts,amsbsy,bm,euscript,mathrsfs}
\usepackage{amssymb,stmaryrd,faktor,slashed}
\usepackage{color}

\DeclareFontFamily{OT1}{rsfs10}{}
\DeclareFontShape{OT1}{rsfs10}{m}{n}{ <-> rsfs10 }{}
\DeclareMathAlphabet{\mathscript}{OT1}{rsfs10}{m}{n}
\numberwithin{equation}{section}

\newcommand{\tr}{\text{tr}}

\def\a{\alpha}
\def\b{\beta}
\def\g{\gamma}

\def\k{\kappa}
\def\l{\lambda}
\def\m{\mu}

\def\s{\sigma}

\def\F{\Phi}
\def\G{\Gamma}

\def\gsim{ \lower .75ex \hbox{$\sim$} \llap{\raise .27ex \hbox{$>$}} }
\def\lsim{ \lower .75ex \hbox{$\sim$} \llap{\raise .27ex \hbox{$<$}} }
\def\be{\begin{equation}}
\def\ee{\end{equation}}
\def\bea{\begin{eqnarray}}
\def\eea{\end{eqnarray}}

\def \td {\tilde}

\def \V  {{\rm V}}
\def \ha {{1 \ov 2}}

\def \sql {{\sqrt{\l}}\ }
\def \del{\partial}
\def \a {\alpha}
\def \aa {{\a'}}
\def\ov{\over}
\def \ci {\cite}

\def \foot {\footnote}
\def \bi{\bibitem}
\def\la{\label}\def\foot{\footnote}\newcommand{\rf}[1]{(\ref{#1})}

\def \no {\nonumber}
\def \adss {$AdS_5 \times S^5\ $}
\def \a {\alpha }

\def \V  {{\rm V}}
\def\ket {\big\rangle}
\def\bra {\big\langle }

\def \N  {{\cal N}}
\def \adt  {{$AdS_2\times S^2$\ }}

\def \vee {\ep}

\def \ed {\end{document}}

\begin{document}
\begin{titlepage}
\vspace{-1cm}
\vspace{-5cm}
\title{
  \hfill{\small Imperial-TP-AT-2015-07  }  \\[2 em]
   {\LARGE   Heat  kernels on  cone of $AdS_2$
       and  $k$-wound\\  circular Wilson  loop 
    in $AdS_5 \times S^5$ superstring 
    }
\\[1em] }
\author{
   R. Bergamin
     \\[0.1em]
   {\it \small Dipartimento di Fisica e Astronomia  ``Galileo Galilei" }\\
       {\it \small Universit\`a degli studi di Padova,  Via Marzolo 8, 35131 Padova, Italy}\\[0.5em]
A. A. Tseytlin\footnote{Also at Lebedev Institute, Moscow. tseytlin@imperial.ac.uk}
\\[0.1em]
   {\it \small The Blackett Laboratory, Imperial College, London SW7 2AZ, U.K.}
   }
\date{}
\maketitle
\begin{abstract}
We compute  the one-loop correction to   partition function  of $AdS_5 \times S^5$   superstring 
that should be representing $k$-fundamental   circular Wilson loop in planar limit. 
The  2d  metric of the minimal surface ending on $k$-wound  circle  at the boundary 
  is that of a cone  of $AdS_2$  with deficit $2\pi (1-k)$.  We compute the determinants of 2d   fluctuation operators  by 
first  constructing  heat kernels of scalar and spinor Laplacians   on  the  cone 
using Sommerfeld formula.  The final  expression   for the $k$-dependent part   of the one-loop 
correction  has simple integral representation  but is   different from earlier results. 
\end{abstract}

\thispagestyle{empty}

\end{titlepage}
\def \iffa {\iffalse}
\def \tr  {{\rm tr\,}}
\def \F {{\cal F}} 
\def \smm {{\rm sym} }  \def \edd {\end{document}}
\def \N {{\cal N}}
\def \te {\textstyle}
\def \ha  {{\te {1\ov 2}} }
\def \be {\bea}
\def \ee {\eea} 
\def \aa  {{\rm a}}
\def \ep   {\epsilon} 
\def \ad {AdS_2 \times S^2} 
\def \be {\bea}
\def \ee {\eea}
\def \cc {{\rm c}} 
\def \KK {{\rm K}} 
\def \lm {v} 
\def \tK {\td K} 
\def \mm {{\rm m}}
\def \ss {{\rm s}}  \def \ff {{\rm f}} \def \gv {{\rm g}} 
\def \vv {{\rm v} }
\def \ddd {d} 
\def \adt {{$AdS_2$} }
\def \tG  {{\G}} 

\def \hG {{\hat \G}}

\def \Tr { \mathrm{Tr\, }}
\def \hal {{1\ov 2}}
\def \KKF {\KK_{_\hal}} 
\def \KF  {K_{_\hal}} 
\def \FF {F_{_\hal}}
\def \fo {{\te { 1 \ov 4}}}

\def \sadt {_{AdS_{2}} }
\def \adk  {$AdS_{2\k}\ $} \def \fo {{1\ov 4}}

\tableofcontents
\section{Introduction  and summary}\label{secint}

In    this paper, which is a sequel to Appendix B in  \ci{bt},   we shall  consider the  computation of 
the disc     partition function  for  the \adss   superstring
\be \la{0}  Z=   e^{- \tG}  \ , \ \ \ \ \ \qquad     \tG= \tG_0 + \tG_1 + ...  \ee
 for  the euclidean minimal surface  of  ending on $k$-wound   circle at the boundary. 
 Here the expansion is in inverse  powers of the string tension $T= { \sql\ov 2\pi}$  and we will   be interested in  first subleading (1-loop)  correction $\tG_1$. 
  
The string  partition function  should be representing  \ci{ber,dgo}    the expectation value $\langle W \rangle$ of the circular  Wilson loop  
in $k$-fundamental $SU(N)$  representation  in dual    $\N=4$ SYM theory 
 taken   in the limit   $N\to \infty$ and then expanded in  large `t Hooft coupling 
 $\l$ for fixed $k$. The well-known exact  expression for $\langle W \rangle$ in the planar approximation  gives 
 \ci{Erickson:2000af,Drukker:2000rr,Drukker:2005kx,Pestun:2007rz}
\bea
&& \langle W \rangle ={ 2  \ov  k \sql }  {I}_1 ( k \sql)=  e^{- \hG} \ , \ \ \ \ \ \qquad      \hG= \hG_0 + \hG_1 + ... 
 \ , \la{1} 
\\
&& \hG_0 = - k \sql  \ , \qquad \qquad  
\te \hG_1 =   \ha \ln { \pi \ov 2}  + { 3 \ov 2} \ln k + { 3 \ov 2} \ln \sql 
\ .  \la{b3} \eea
Assuming that the  planar $k$-fundamental    circular Wilson loop 
 is indeed represented by the open 
string path integral  in \adss  with  the boundary condition that the disc-like  world sheet  ends on 
$k$-wound circle at the boundary  the corresponding minimal surface in, e.g.,  
$AdS_3$  with the metric $ds^2 = z^{-2} ( dr^2 + r^2 d \phi^2 + dz^2 )$
is  described  by 
\be \la{2}
 \phi = k \tau\ , \ \ \ \ \ z= \tanh   k \s\ , \ \ \ \ \ 
 r= \cosh^2 k \s \ , \ \ \ \ \ \ \  z= \sqrt{1- r^2} \ , \ee
  where $\tau \in (0, 2 \pi)\ , \ \s\in (0, \infty)$. Then the   induced  metric  is 
\be \la{3} 
ds^2 =  { k^2 \ov  \sinh^2 k\s} ( d\tau^2 + d \s^2) = d \xi^2 + k^2  \sinh^2 \xi\,  d \tau^2 \ , \qquad \qquad 
\te \xi =\ln  \tanh {k\s\ov 2}\ . 
\ee
 For $k=1$   this is  the usual   $AdS_2$ metric   but  for  $k=2, 3, 4, ...,$
  it has  a conical singularity   at $\xi=0$   with negative deficit $\delta= 2 \pi (1-k)$, i.e. this is a cone of $AdS_2$.  
The   corresponding    classical action is proportional to  the regularized $AdS_2$  volume.
To recall,  the euclidean  ${\widetilde{AdS}_2}$  space in Poincare coordinates   $ds^2 = z^{-2} ( dx^2  + dz^2 )$   with  boundary  being a line $R$ 
 is 
not   equivalent to $H_2= AdS_2$   in  angular  coordinates $ds^2=  d \xi^2 +  \sinh^2 \xi\,  d \varphi^2$ 
    with  boundary $S^1$; in particular,  
 their  regularized   volumes are different 
\be  V_{{\widetilde{AdS}_2} }= { \ell   \ov \vee} \to\  0 \ , \ \ \ \ \ \ \ \ \ \ \ \ \ 
 V_{{{AdS}_2} }=2 \pi  ( { 1 \ov \vee} - 1)   \to \  - 2 \pi  \ .   \la{o}\ee
Here  we set the  $AdS_2$ scale  to 1,     $\ell \to \infty$ is the length of the boundary of ${{\widetilde{AdS}_2} }$
  and  $\vee\to 0$ is a radial  IR cutoff.
As a result, the renormalized    classical string action  for \rf{3}    is 
$\tG_0= - k\sql$    which matches the value of $\hG_0$ in \rf{b3}. This  suggests that the prescription 
to use the regularized   volume of $AdS_2$ is a required part of definition of string theory  partition function    in the context of AdS/CFT.

The starting point to  compute  $\tG_1$ 
is the  general form of the  \adss superstring one-loop   correction 
  \cite{dgt,thei}\foot{Here $\nabla^2_0$
and $\nabla^2_\hal$ are 2d Laplacians   defined on scalars and  spinors respectively.
$R$ is the curvature of 2d metric equal to $-2$ in the case of $AdS_2$ of unit radius.  }
\be 
\tG_{1}=  \ha  \ln \frac{[{\rm det} (-\nabla^2_0 )]^5 \, [{\rm det} (-\nabla^2_0 +2  )]^3 \ }{[{\rm det} (-\nabla^2_{{1\ov 2}} +\frac{1}{4}R +1) ]^8}\,. 
\label{b1}
\ee
In the simplest  cases   of  minimal surfaces  with  symmetric-space 
  induced metric  (like $AdS_2$   spaces  of straight  line and circular  loop) it 
 is  natural to preserve this symmetry  in the loop  computation till very end, 
i.e.  not to introduce radial cutoff  directly  on contributions of individual modes \ci{dgt,bt}. Then, as standard    for determinants on a symmetric spaces
\ci{cha,cam,Fursaev:2011zz} 
 they can be computed using heat kernel technique  (utilizing in an   essential  way  the unbroken global symmetry) 
 and   the final expression   will be proportional to the   space  volume  factor for which we should   use its renormalized value in \rf{o}.
 For example, 
 in  the straight  line  case the classical action and loop corrections  will  be  proportional to the volume 
of $\widetilde{AdS}_2$   with   boundary  $R$  which has zero renormalized value in  \rf{o}  and     should  thus be  assumed to   vanish   \ci{dgt,kt},  
 in agreement with $\langle W \rangle =1$ on the gauge theory side.  

Such heat   kernel  based  computation  of \rf{b1}   in the $k=1$  circular loop  case  gives (using that 
the renormalized  $V_{{{AdS}_2} }= - 2 \pi$)
 \ci{dgt,bt} 
\be  \tG_{1} (1)  =   \ha \ln (2 \pi) \ .   \la{4} \ee
The same  expression was found in \ci{kt}  using  a different (Gelfand-Yaglom) method to  compute 
the determinants  with  the  radial IR    cutoff   implemented explicitly on modes 
and all power  divergences dropped in the final expression. The same  method applied  to the  $k>1$ case   led to the following 
generalization of \rf{4}  \ci{kt}
\be 
\tG_{ 1\,   {\rm KT} } (k) =   \ha \ln (2 \pi)  +  ( 2 k + \ha) \ln k  - \ln  k!   \ .   \la{b2} \ee
This expression  does not match the   gauge theory result $\hG_1$ in \rf{b3}. 
The third $\ln \sql$  term  in $\hG_1$  may be attributed to   the presence of the  string tension 
 normalization factor for  the three  (Mobius-symmetry) 
  ghost zero modes on the disc \cite{Drukker:2000rr}, which was 
not included in \rf{b1} and thus should be absent in  \rf{4},\rf{b2}.\foot{Note that   this  contribution   should be absent in  the straight line  case 
as the  induced metric volume enters also  the normalization of the zero modes   but  the renormalized   $V_{\widetilde{AdS}_2}=0$   
according to \rf{o}.}
The remaining     difference    may be coming from  a numerical  factor  in   normalization of  the  disc zero modes 
or   from the ratio of the ghost and  the two longitudinal mode determinants  (assumed \ci{dgt} to be  equal to  one 
 in \rf{b1}) once they are  computed   with proper boundary 
conditions (cf.  \ci{km}).

\def \II  {{\rm I}}

Below  we shall ignore these  open issues   and 
 aim  at  just   deriving   the analog of \rf{b2}   in the  heat kernel   approach 
of  \ci{dgt,bt}. Explicitly, we will find   that 
\bea
 \tG_1 (k) =  \ha k \ln (2 \pi)   +    \II(k) \ , \ \ \ \qquad  \ \ \ \ \    \II(1) =0  \ ,  \qquad \qquad  \ \ 
 \la{55} \\
  \la{5} \II(k) = -  \frac{1}{4} \int_{0}^{\infty} {
  dy\ov y \, \sinh y} \Big[  (  5 e^{-y} + 3 e^{-3y} ) \big( \coth \frac{y}{k}-k\coth y \big) 
  +  16   e^{-2y}  \big( { 1 \ov \sinh { y \ov k} }  - { k  \ov \sinh y} \big) \Big] 
\ee
Here  the  terms with $  e^{-y} $ and $ e^{-3y}$   are  the contributions of  the scalar   factors under log  in \rf{b1} 
and  the  term with $  e^{-2y} $   -- the contribution  of the   spinor  factor. 
For large $k$   the integral $I(k)$  is well approximated by a  straight line, $\II(k) = c\,   k  + ..., \  c=1.077$. 
This  result  for  $\tG_1$  with  $k >1$   is   thus  different  from \rf{b2} of \ci{kt}
and also different from  the simple   gauge-theory   expression $\sim \ln k$ in \rf{b3}.

Here we will   not  attempt to understand the difference with \rf{b2}, 
concentrating    on developing 
 technical  tools 
  required   for derivation of  \rf{5}  that may have other potential applications. 
The main issue    is   to find  scalar and fermion heat kernels  on the  cone of \adt  in \rf{3}. 
Let us mention  also   that another possible   application of  the above  partition function  
is to computation of  R\'enyi  entropy on $AdS_2$  (see \ci{sol}).

It  will be  useful  to   replace  $k$  in \rf{3}  by  an arbitrary real number $\k$  and  set $\k=k$ only at the end. 
Then if  we    take  $\k$    to be  equal to 
 $1/n$  with an integer $n$   the  metric  \rf{3}   becomes  that of 
 an orbifold $AdS_2/Z_n$ (with a conical singularity of positive deficit) 
  and the corresponding  scalar  or fermion 
heat kernel  can be found  as a sum over images  \ci{ms,jon,gul}.
This  special case of  $\k = 1/n$   will thus   allow us  to check the general $\k$ expressions 
for heat kernels. 

In the case of   a  cone of a 2-plane  $ds^2 = d \xi^2 + \k^2  \xi^2   d \tau^2$   with a 
 general   $\k$   the  scalar   Laplacian  heat 
  kernel can be found using ``re-periodisation"  trick  (or Sommerfeld formula) 
\ci{dowk} and the same idea applies  also   to the   construction  of 
  heat  kernel on the cone of $AdS_2$  with generic $\k$    \ci{ms}. 
We shall   present  a detailed  derivation of  the expression  of \ci{ms} for 
the  scalar heat kernel  in section 2. 

The   analogous   construction   for  the  spinor   heat kernel  can be given 
 using  the result  \ci{camp,camf} for the regular $AdS_2$ 
and generalizing  the   examples  of  cones of  2-plane and 2-sphere in  \ci{kab,fm}. 
We shall present it in section 3.  This will be our main  new  technical  result. 
As a check, we shall  match the   small $t$ expansion of the spinor kernel   with the general  form of  asymptotic 
expansion of spinor heat kernel on a conical singularity in \ci{furs}   and also show that 
 in the special  case  of $\k=1/n$ corresponding to  $AdS_2/Z_n$  orbifold  it  agrees  with the
 result in  \ci{gul}. 
 
 In section 4   we shall   use the results for  the scalar and spinor heat kernels to compute the determinants in \rf{b1}
 deriving \rf{5}  and comparing it to \rf{b2}. 
 
In Appendix {A}  we shall review  the derivation of the Sommerfeld formula 
for  heat kernels  of   2d scalar and spinor Laplacians 
on a manifold with a conical singularity.
In Appendix {B} we shall present the derivation of the  phase factor matrix $U$ in the \adt 
spinor heat kernel in \rf{31}.

\section{Scalar   heat  kernel on  cone of $AdS_2$  } 
In order to compute the $1$-loop  partition function \rf{0},\rf{b1} we need to know the
traces  of  heat kernels 
of the scalar and spinor Laplacians
 on the  cone  of \adt  with an arbitrary  deficit $ \delta= 2 \pi (1-\k)$ (we shall  denote this space as  $AdS_{2\k}$).
 In the regular \adt case we have 
\bea \la{ab1} 
 \ln \det \Delta = -    \int_\ep^\infty { dt \ov t} \   \KK(t) \ , \ \ \ \ 
\KK(t) = \int d^2 x \sqrt g \  K(x,x;t) \ , \ \ \   K(x,x';t) = \bra x| e^{-t \Delta} | x'\ket 
\ \ \ \\
 \KK(t) =  V_{{AdS_{2}}}  K(t) \ , \ \ \
 \ \ \ \ \ \ \ \ K(t) =   K(x,x; t) \ , \ \ \ \ \ \ \ \ \ \ \  \ \ \ \ 
\  V_{{AdS_{2}}} = - 2\pi  \  . \la{vvo}
\eea
Here $\KK$ is the  trace of heat  kernel which on symmetric space is a  product of (regularized) volume and   heat kernel at coinciding points. 

In  this    section we will review the computation of the heat kernel $\KK$  of the scalar Laplacian 
$\Delta_0= - \nabla^2_0 + m^2$  on  $AdS_{2\k}$
 using the Sommerfeld formula, rederiving the  expression  for it given in  \ci{ms}.
 We shall    check the consistency of  the result  in the special  case when  $\k = 1/n$  with integer $n$  
  by   comparing it with the scalar heat kernel on the orbifold $AdS_{2}/Z_{n}$.
  
 The generalization of the scalar heat kernel on  regular $AdS_{2}$ to the case of  \adk
 can be  constructed  using the  Sommerfeld formula \ci{dowk} as reviewed  in Appendix {A}. 
  The   starting   point is  the expression for the untraced heat kernel of the scalar Laplacian  
  on the regular $AdS_{2}$
  \ci{cha,cam}
\begin{equation}
K_{AdS_{2}}(x,x^{\prime};t)= \frac{\sqrt{2}\, e^{-(m^{2}+\fo)t}}{(4\pi t)^{{3}/{2}}} \int_{d(x,x^{\prime})}^{\infty} \frac{dy \  y \  e^{-\frac{y^{2}}{4t}}}{\sqrt{\cosh y-\cosh d(x,x^{\prime})}}\ . \la{21}
\end{equation}
Here $d(x,x')$  is  the geodesic distance   between  the   two  points $x=(\xi,\phi)$ and $x'=(\xi', \phi')$ 
of \adt   in  polar  coordinates 
\bea &&  ds^2 = d \xi^2 + \sinh^2 \xi \, d \phi^2 \ , \la{aa}\\
&& \cosh d(x,x')
= \cosh \xi\,  \cosh \xi'-   \sinh \xi\,  \sinh \xi'\,  \cos(\phi-\phi')  \ . \la{aaa} \eea
 According to the Sommerfeld formula, the traced heat kernel in presence of a conical singularity when $\phi \in (0, 2 \pi \k)$ 
  is given by\foot{The $AdS_2$ scalar heat kernel has also an alternative representation 
  
  $K_0 (t) =   
   \frac{e^{-(m^{2}+\fo)t}}{2\pi}    \int_{0}^{\infty} dv \, v\,   \tanh( \pi v)   \,   e^{- t v^2} .$}
\bea
&& \label{S}\KK_{AdS_{2\k}}(t) \equiv \KK_{\k}(t)=  K(0; t ) + \frac{i}{4\pi \k} \int_{\Gamma} dz \, \cot  \frac{z}{2\k}\ K(z;t)\ , \\  
&& K(z;t)\equiv  \int_{0}^{2\pi \k}d\phi \int_{0}^{\infty}d\xi \ \sinh \xi \ K(\xi=\xi^{\prime},\phi^{\prime}=\phi+z;t)\ ,\la{ss}  \\
 && K(\xi=\xi^{\prime},\phi^{\prime}=\phi+z;t)=\frac{\sqrt{2}e^{-(m^{2}+\fo)t}}{(4\pi t)^{3/2}} \int_{d(\xi,z)}^{\infty} \frac{dy\, y \, e^{-\frac{y^{2}}{4t}}}{\sqrt{\cosh y-\cosh d(\xi,z)}}\ , \la{sss}\\ 
 && K(0;t)= \int_{0}^{2\pi \k}d\phi \int_{0}^{\infty}d\xi\ \sinh \xi \ K(\xi=\xi^{\prime},\phi^{\prime}=\phi;t)= \k \, \KK_{AdS_{2}} (t) \ , 
 \la{ses}\\
  &&     \KK_{AdS_{2}} (t) =  V_{{AdS_{2}}}  \,  K_0(t)\ , \ \ \ \   \ \ \  K_0(t)  = \frac{e^{-(m^{2}+\fo)t}}{(4\pi t)^{{3}/{2}}} \int_{0}^{\infty} {dy \, y\,   {1 \ov \sinh {y\ov 2}}  \,   e^{-\frac{y^{2}}{4t}}}  \ . 
   \la{sis} \ \ 
 \eea
 Here $K(z;t)$  is the  integrated heat kernel with points   separated   by $z$ along  angular direction only.
 The contour $\Gamma$ on the complex plane  is made of two vertical lines going from
  $ -\pi +i \infty $ to $-\pi - i \infty$ and from $ \pi -i \infty $ to $\pi +i \infty$ (see Appendix {A}). 
   The geodesic distance $d(\xi,z)$ between the points $(\xi,\phi)$ and $(\xi,\phi+z)$ is  given by 
    $\sinh d(\xi,z)= | \sinh \xi\ \sin \frac{z}{2}|$. 
    
    In \rf{sis}   we 
     used that   for  regular   symmetric  $AdS_2$ space the  heat  kernel   does not depend on     coordinates  at coinciding points. 
    Note that the IR divergence regularized by $ V_{{AdS_{2}}}  \to - 2 \pi$    will be present only in the first  
    ``$AdS_2$"  term    $K(0;t)$  in \rf{S}, while the second integral term will be finite.  
 
 To compute $K(z;t)$  we introduce $v= \sinh \xi, 
\  \cosh d(v,z) =1+2v^{2}\sin^{2}\frac{z}{2}$  to get 
\bea
 K(z;t)
=  \frac{\k\ e^{-(m^{2}+\fo)t}}{2(4\pi)^{{1}/{2}}\ t^{3/2}\ \sin {z\ov 2}} \int_{0}^{\infty} \frac{dv\  v }{\sqrt{1+v^{2}}} 
\int_{d(v,z)}^{\infty} 
{dy \ y\  e^{-\frac{y^{2}}{4t}} \ov \g^2(y,z)-v^{2} }\ ,\la{22} \, \ \qquad  \g(y,z)\equiv \frac{\sinh{\frac{y}{2}}}{\sin{\frac{z}{2}}} 
\ .\la{23}
\eea
Now we interchange the integration limits 
$ \int_{0}^{\infty}dv \int_{d(v,z)}^{\infty} dy =
\int_{0}^{\infty}dy \int_{0}^{d^{-1}(y,z)  }dv, \ \   
$
and integrate over $v$ using 
$
 \int_{0}^{\g(y,z)} \frac{dv\ v}{\sqrt{1+v^{2}} \sqrt{\g^{2}(y,z)-v^{2}}}= \frac{\pi}{2}-\arctan \g^{-1} (y, z) 
$.
Then integrating by parts one  finds  that 
\begin{equation}\la{24}
K(z;t)= 
\frac{\k}{(4\pi t)^{{1}/{2}}\sin {z \ov 2} } \int_{0}^{\infty}
dy\ e^{-{y^{2}\ov 4t}} \ \frac{\cosh {y\ov 2} \ \sin {z\ov 2} }{ \cosh y-\cos z} \ . 
\end{equation}
Changing the variable $ y\rightarrow 2y$ we  finally get 
\begin{equation}
\la{25}
K(z;t)=
\frac{\k \ e^{-(m^{2}+\fo)t}}{(4\pi t)^{{1}/{2}}} \int_{0}^{\infty}
dy \ e^{-{ y^2 \ov t} } \  \frac{\cosh y}{\sinh^{2} y + \sin^{2} { z \ov 2} }
\ . 
\end{equation}
According to (\ref{S}), the scalar heat kernel on  cone of $AdS_{2}$  is thus given by 
\bea
&& \KK_{\k}(t)= \k\,  \KK_{AdS_{2}}(t) + \frac{\k\ e^{-(m^{2}+\fo)t}}{(4\pi t)^{{1}/{2}}} \int_{0}^{\infty}
dy\ \cosh y\ e^{-{y^{2}\ov t} } \  \frac{i}{4\pi \k} \int_{\Gamma} dz\ F(z;y,\k)\ , \la{26} \\ 
&& F(z;y,\k)=\frac{1}{\sinh ^{2} y+\sin ^{2}{z \ov 2} }\  \cot \frac{z}{2 \k}\ . \la{27} 
\eea 
For $\k=1$  the function  $F(z,y;\k)$ becomes $2 \pi$ periodic in $z$, and  
due to the structure of the contour $\Gamma$, this leads to the  vanishing of the integral contribution. 

To evaluate the contour integral we  use  the residue  theorem.
 The two vertical lines $ (-\pi +i \infty ,-\pi - i \infty)$ and $ (\pi -i \infty ,\pi +i \infty)$  surround 
  one pole of $ \cot \frac{z}{2\k}$ at $z=0$ with the residue
$
\mathrm{Res} \ F(z=0;y,\k)= \frac{2\k}{\sinh^{2} y}
$.
Setting  $ y=-iu$  we get $ F(z;u,\k)= \frac{\cot \frac{z}{2 \k}}{ (\sin  \frac{z}{2 } +\sin u)(\sin    \frac{z}{2}-\sin u)}$
  with  two poles on the imaginary axis at $ z= \pm 2u $. 
Using the limit
\begin{equation}
\lim_{z\rightarrow \pm 2u} \frac{\sin  \frac{z}{2}  \pm \sin u}{z \pm 2u}= \ha \cos u 
\end{equation}
we find
\begin{equation}
\mathrm{Res}\ F(z= \pm 2u;u,\k)=\frac{\cot \frac{u}{\k}}{\sin u \cos u}= -\frac{\coth \frac{y}{\k}}{\sinh y \cosh y}.
\end{equation}
Therefore 
\begin{equation}
\frac{i}{4\pi \k} \int_{\Gamma} dz\,  F(z;y,\k)=2 \pi i \ \frac{i}{4\pi \k} 
 \Big( \frac{2\k}{\sinh^{2}y} -\frac{2 \coth \frac{y}{\k}}{\sinh y \  \cosh y}\Big) = \frac{1}{\sinh^{2}y}\Big( \frac{\tanh y}{\k \tanh \frac{y}{\k}}-1\Big) \ . \la{277}
\end{equation}
Finally,   the  scalar heat kernel on  $AdS_{2\k}$  is given by 
\begin{equation} \label{28}
\KK_{\k}(t)= \k\,  \KK_{AdS_{2}} (t) + \frac{  e^{-(m^{2}+\fo)t} }{(4\pi t)^{{1}/{2}}}\int _{0}^{\infty} dy \ e^{-\frac{y^{2}}{t}}\frac{1}{\sinh y}
\big( 
\coth \frac{y}{\k}-\k\, \coth y \big)  \ , 
\end{equation}
 which is the same expression as  given  in \ci{ms}.
 
The integral vanishes for $\k=1$, i.e.    $ \KK_{\k=1}(t)= \KK_{AdS_{2}}(t) $.
It is   obviously convergent  at $y \to \infty$    and  also  at   $y\rightarrow 0$  since 
$\frac{ \coth \frac{y}{\k} -\k \coth y  }{\sinh y} \big|_{y\to 0}   \to 
\frac{1}{3} ({\k}^{-1} - \k ) + O(y) $.

The special  case  of  $\k = 1/n$     corresponds to the 
 $Z_{n}$ orbifold of $AdS_{2}$  which is   obtained by identification 
 $
\phi = \phi + \frac{2 \pi}{n} $.
The  analog  of the Sommerfeld formula for orbifolds   is  equivalent to the  summation over images expression 
 discussed  in \cite{gul}. There  it was  found  that the small $t$ expansion of the heat kernel for a massless scalar
  Laplacian on $ AdS_{2}/Z_{n}$ is given by 
 \begin{equation}\label{30}
\KK_{AdS_{2}/Z_{n}}(t)=\te  \frac{1}{n}  \KK_{AdS_{2}}(t)+ \frac{n^{2}-1}{12n} -
\frac{(n^{2}+11)(n^{2}-1)}{360n}\, t   +O(t^{2}) \ .
 \end{equation}
 To compare to the general expression \rf{28} we need first  to set there $m=0$  and expand  for small $t$. 
  Changing the  variable $y\rightarrow  \sqrt{t}\, u$ we get for $t\to 0$ 
\begin{equation}\la{31a} 
\begin{split}
\KK_{\k}(t)= & \k\,  \KK_{AdS_{2}}(t) + \frac{\k ( 1-\frac{t}{4})}{(4\pi)^{{1/2}}} \int _{0}^{\infty} du\   e^{-u^{2}}  \ \te  \big[ \frac{1}{3} \big(\frac{1}{\k ^{2}}-1\big) +  \frac{7\k^{4}-5 \k^{2}-2}{90 \k^{4}} \, t \, u^{2} \big] + O(t^{2})\\ 
=& 
\k\,  \KK_{AdS_{2}}(t)+ \te   \frac{1}{12}\big(\k^{-1} - \k \big) -  \frac{1+10\k^{2}-11\k^{4}}{360 \k^{3}}\, t  + O(t^{2})\ . 
\end{split}
\end{equation}
Setting here $\k= 1/n$ we  indeed match the orbifold expansion \rf{30}.

\section{Spinor    heat  kernel on  cone of $AdS_2$  }

Let us now turn to the  spinor  case.  
The fermions  in \rf{b1} are  real (Majorana)  so that log of their determinant   or trace of heat kernel  should be understood with extra $\ha$  included
(so that on a trivial background  the contribution of one 2d Majorana  fermion  is minus that of one real scalar). 
 The  starting point is the   heat kernel for the (Majorana) spinor Laplacian  $\Delta_\hal= - \nabla^2_\hal  + {1 \ov 4} R + m^2 $ 
 on $AdS_{2}$    \ci{camp} 
\begin{equation} \la{31} 
K(x,x^{\prime};t)= U(x,x^{\prime})\, \frac{
e^{-m^{2}t}}{\sqrt 2 \, (4\pi t)^{3/2}} { 1 \ov \cosh {d(x,x^{\prime})\ov 2} } \int_{d(x,x^{\prime})}^{\infty} \frac{dy\,  y\, \cosh \frac{y}{2} \, e^{-\frac{y^{2}}{4t}}}{\sqrt{\cosh y-\cosh d(x,x^{\prime})}}\ .
\end{equation}
Here $U(x,x^{\prime})$  is a $2\times 2$ matrix acting on the $AdS_{2}$ 
spinor bundle and satisfying the equation of the parallel transport along the geodesic connecting the two points $(x,x^{\prime})$:
\begin{equation}\label{32}
n^{\m}(x,x^{\prime})\nabla_{\m} U(x,x^{\prime})=0 \ , \qquad 
U(x,x)= I  \ , \qquad  n_{\m}(x,x^{\prime})=\nabla_{\m}\, d(x,x^{\prime})= \partial_{\m}\,  d(x,x^{\prime}) \ . 
\end{equation}
Here the  derivatives are taken with respect to the point  $ x $  and $ n_{\m}(x,x^{\prime})$ 
 is the tangent vector to the geodesic. 
The general structure of  $U(x,x')$  on hyperbolic spaces was not spelled out in the literature. 
As we show in  Appendix {B} in the present case of  $AdS_2$ in polar  coordinates \rf{aa}   it is given by 
($\s_3$  is the Pauli matrix) 
\begin{equation}\la{4u}
U(\xi,\xi^{\prime},\phi,\phi^{\prime})= \exp\Big(  i \sigma_{3}\,  \arctan \frac{\cosh \frac{\xi+\xi^{\prime}}{2}\, \tan\frac{\phi-\phi^{\prime}}{2}}{\cosh \frac{\xi-\xi^{\prime}}{2}} \Big)\  .
\end{equation}
We can now 
apply the fermionic version of the Sommerfeld formula (see Appendix {A})
to obtain the heat kernel of the spinor Laplacian on the  cone   of $AdS_{2}$ (cf. \rf{S},\rf{ss})
\bea
&&\la{F} \KK_{\k}(t)= \Tr K(0;t )+ \frac{i}{4\pi \k} \int_{\Gamma} dz\ \frac{1}{\sin  \frac{z}{2\k}}\ \mathrm{Tr\, }  K(z;t)\ ,\\
&& \la{ff}
\mathrm{Tr\, } 
 K(z;t)\equiv  \int_{0}^{2\pi \k}d\phi \int_{0}^{\infty}d\xi\,  \sinh \xi \ \mathrm{Tr\, }K(\xi=\xi^{\prime},\phi^{\prime}=\phi+z;t)\ , \\
 &&
  \mathrm{Tr\, }K(\xi=\xi^{\prime},\phi^{\prime}=\phi+z;t)=  \mathrm{Tr\, }U(\xi,z)\, 
  \frac{
  e^{-m^{2}t}}{\sqrt 2\, (4\pi t)^{3/2}}{ 1 \ov \cosh {d(\xi,z)\ov 2}} 
  \int_{d(\xi,z)}^{\infty} \frac{dy\,  y\, \cosh \frac{y}{2} \,  e^{-\frac{y^{2}}{4t}}}{\sqrt{\cosh y-\cosh d(\xi,z)}} \no\\
 && U(\xi,z)= \exp \Big[  i \sigma_{3} \arctan \big(  \cosh \xi \tan\frac{z}{2}\big) \big]\ , 
 \qquad \  \Tr U(\xi,z)=\frac{2}{\sqrt{1+\cosh ^{2} \xi\ \tan ^{2} {z\ov 2}  }}\ , \la{fff}\\
  &&\la{FA}
 \Tr K(0;t)= \Tr K(x,x;t) =   \k\,  \KK_{AdS_2} (t) \ , \ \ \qquad     \KK_{AdS_2}  (t) = V_{AdS_2} \KF(t) \ , \\
&& \la{Fv} 
 \KF (t) = \frac{
 e^{-m^{2}t}}{(4\pi t)^{3/2}}  \int_{0}^{\infty} {dy\,  y\, \coth \frac{y}{2} \, e^{-\frac{y^{2}}{4t}}}    \ .
\eea
Here $K(x,x';t)=K(\xi,\xi',\phi,\phi';t)$  is  the fermionic kernel in  the regular \adt case in \rf{31}.\foot{The trace of the 
$AdS_2$  Majorana spinor
 heat kernel has also an alternative representation
  
  $\KF (t) =   
   \frac{e^{- m^{2}t}}{2\pi}    \int_{0}^{\infty} dv \, v\,   \coth( \pi v)   \,   e^{- t v^2} .
   $

   \noindent
    Note that in contrast to \ci{bt} here 
 we do  not include  the fermionic statistics minus  sign in the definition of the fermionic 
  heat kernel,  accounting for  this   sign explicitly  when combining the bosonic and  fermionic contributions below.
   } 
The contour $\Gamma$ is the same as in  the scalar   case in \rf{S} (see Appendix {A}).

To compute $\Tr K(z;t)$   in \rf{ff} we  integrate over  the angle   and 
 after a  change of variables 
 ($v = \sinh \xi, \ \ 
   \cosh d=1+2v^{2}\sin^{2}\frac{z}{2}$)  we  find 
\begin{equation}\la{33}
\Tr K(z;t)=\frac{\k \, e^{-m^{2}t}}{(4\pi)^{{1}/{2}}\, t^{3/2}} \int_{0}^{\infty} 
\frac{dv\, v }{\sqrt{1+v^{2}}} \frac{\cos {z\ov 2}}{1+v^{2}\sin^{2} {z\ov 2} }\int_{d(v,z)}^{\infty} 
\frac{dy\, y\, \cosh \frac{y}{2}\,  e^{-\frac{y^{2}}{4t}}}{\sqrt{\sinh^{2} \frac{y}{2}-
v^{2}\sin^{2} \frac{z}{2}}}.
\end{equation}
Interchanging the integration limits as explained  below \rf{23} 
 we get 
\begin{equation}
\begin{split} \la{355}
&\mathrm{Tr}\,  K(z;t)=\frac{\k\,  e^{-m^{2}t}}{(4\pi)^{{1}/{2}}\, t^{3/2} \sin {z\ov 2} } \int_{0}^{\infty}
dy\,  y\,  \cosh \frac{y}{2}\,  e^{-{y^{2}\ov 4t} } \ G(y,z), \\
& G(y,z)\equiv \int_{0}^{\g(y,z)}{ dv\ov \sqrt{1+v^{2}}} \,  \frac{v \ \cos {z\ov 2}}{ \sqrt{\g^{2}(y,z) -v^{2}}\ (1+v^{2}\sin^{2}{z\ov 2} )}\ , 
\end{split}
\end{equation}
where   $\g(y,z)={ \sinh \frac{y}{2}\ov \sin \frac{z}{2}}$  as in  \rf{23}.
As a result
\begin{equation}
\la{34}
G(y,z)= 
 \frac{1}{\cosh{ y\ov 2} } \big( {\pi \ov 2} -\arctan   \frac{\tan {z\ov 2}  }{\tanh {y \ov 2} }\big)  \ .   
\end{equation}
Integrating by parts in the remaining integral over $y$ in \rf{355} one   can put it into the form 
\bea
\mathrm{Tr}\,  K(z;t)
=  \frac{\k\,  e^{-m^{2}t}}{(4\pi t)^{{1/2}}\sin{z\ov 2}} \int_{0}^{\infty}
dy \ e^{-{y^{2}\ov 4t}}\  \frac{\sin z}{ \cosh y-\cos z }\ . \la{36} 
\eea
Changing the variable $ y\rightarrow 2y$ gives 
\begin{equation}\la{37}
\mathrm{Tr}\,  K(z;t)=   \frac{2\k\,  e^{-m^{2}t}}{(4\pi t)^{{1/2}}} \int_{0}^{\infty}
dy\ e^{-{y^{2}\ov t}} \ \frac{\cos {z\ov 2}}{ \sinh^{2} y + \sin^{2} {z\ov 2}} \ , 
\end{equation}
so that \rf{F} takes the form  (cf. \rf{26}) 
\bea
&& \KK_{\k}(t)= \k\,  \KK_{AdS_2}(t) + \frac{2\k \,  e^{-m^{2}t}}{(4\pi t)^{{1/2}}} \int_{0}^{\infty}
dy\,  e^{-{y^{2}\ov t}}  \ \frac{i}{4\pi \k} \int_{\Gamma} dz\ \FF(z;y,\k)\ ,  \la{38}  \\
&& \FF (z;y,\k)=\ \frac{1}{ \sinh ^{2}y+\sin ^{2} {z\ov 2}}\  \frac{\cos \frac{z}{2}}{\sin {z\ov 2\k}}  \ . \la{39} 
\eea
For $\k=1$ the function  $\FF(z,y;\k)$   becomes  the same as $F(z,y;\k) $ in \rf{27}
  in the scalar case,   so that  the  second integral  term 
 vanishes because of the structure of the contour $\Gamma$.
 
 It remains to evaluate the contour integral using the  residue  theorem. 
  As in the scalar  case,
 the contour $\Gamma$ contains one pole of $ \frac{1}{\sin {z\ov 2\k}}$ at $z=0$ and two imaginary poles $z= \pm 2iy $ 
 of $ \frac{1}{\sinh ^{2}y+\sin ^{2} {z\ov 2}}$, i.e.   (cf. \rf{277}) 
\bea  
 \frac{i}{4\pi \k} \int_{\Gamma} dz\ \FF(z;y,\k)= 
 - \frac{1}{\sinh y}\big(\frac{1}{\k \sinh \frac{y}{\k}}-\frac{1}{\sinh y}\big) \ .  \la{41} 
\eea
Finally,   the spinor heat kernel the $AdS_{2\k}$ is  given by 
\begin{equation} 
\KK_{\k}(t)= 
\k\,  \KK_{AdS_2} (t) {-}  \frac{2\,  e^{-m^{2}t} }{(4\pi t)^{{1/2}}}\int _{0}^{\infty} dy\  e^{-\frac{y^{2}}{t}}\ \frac{1}{\sinh y }\, \big(\frac{1}{ \sinh \frac{y}{\k}}-\frac{\k}{\sinh y}\big) \ . \la{42} 
 \end{equation}
 The integral  term here    vanishes at $\k=1$ and is  convergent at both $y\to \infty$ and $y \to 0$ where 
 $\frac{1}{\sinh y} \big(\frac{1}{\k \sinh \frac{y}{\k}}-\frac{1}{\sinh y}\big) \to
\frac{1}{y}\big[ \big( \frac{1}{y}-\frac{y}{6 \k^{2}}\big) -\big( \frac{1}{y}-\frac{y}{6}\big) \big] =\frac{1}{6} \big( 1-\frac{1}{\k^{2}}\big).
$
 
 To check \rf{42} we may compare its small $t$ expansion with the  asymptotic expansion  of spinor heat kernel 
 in the case  of  a  general curved manifold with a conical singularity derived in \ci{furs}.
 Setting $m=0$ in \rf{42}    and changing $y\rightarrow \sqrt{t}\, u$   we get 
\begin{equation}\la{44} 
\KK_{\k}(t)= \k\,  \KK_{AdS_2}(t) {-}  \frac{2 \k }{(4\pi)^{{1/2}}}\int _{0}^{\infty} du \   e^{-u^{2}}\ \frac{1}{\sinh( \sqrt{t}\, u)}\ \big[\frac{1}{\k \sinh(\frac{\sqrt{t}\, u}{\k})}-\frac{1}{\sinh(\sqrt{t}\, u)}\big] \ , 
\end{equation} 
so that  expanding  in small $t$    gives 
\begin{equation}\la{45} 
\KK_{\k}(t)=  \k\,  \KK_{AdS_2} (t) \te +  \frac{1}{12}   ({\k ^{-1}}-\k)   +  \frac{17\k^{4}-10\k^{2}-7}{1440 \k^{3}}\, t+O(t^{2}) \ , 
\end{equation}
in agreement with general  expressions in  \ci{furs}. 
 As  in the scalar case,   we may  also  match   \rf{45}    with the  small $t$ expansion    
 of the  the massless spinor heat kernel on $ AdS_{2}/Z_{n}$ in \ci{gul} corresponding 
  to  the special case of    $\k= {1\ov n}$
  (cf. \rf{30})
 \begin{equation}\label{43}
\KK_{AdS_{2}/Z_{n}}(t)= \te \frac{1}{n}\KK_{AdS_2} (t)  +  \frac{n^{2}-1}{12n} -
\frac{(7n^{2}+17)(n^{2}-1)}{1440n}\, t  +O(t^{2}) \ . 
 \end{equation}

\section{One-loop partition function}

Let us now apply the   above results for heat kernels \rf{28} and \rf{42}  
 to compute 
the \adss   string partition function \rf{0},\rf{b1}   on  the cone  $AdS_{2\k}$. 
Using the explicit  values of masses ($m^2_0= 0, 2$ and $m^2_\hal =1$)  in \rf{b1} 
and \rf{ab1} we get 
\bea  &&\la{51}
\tG_{1} (\k) = \k\,  \tG_1(1)    + \II(\k)   \ ,  \qquad \   \ \ \  \ \ \ \ \tG_1(1) = \ha \ln (2 \pi) \ ,\  \\
&&  \II(\k) = 
-  \ha  \int^\infty_\varepsilon   { dt \ov t}  \int _{0}^{\infty} {dy}  \ { e^{-\frac{y^{2}}{t}} \ov (4\pi t)^{{1}/{2}}  }\  H(t,y; \k) 
\ , \la{522}  \\
&&  \la{52}  
H(t,y; \k)=   {1\ov \sinh y} \Big[
 ( 5 e^{- { 1\ov 4} t }   + 3 e^{- {9 \ov 4}t  }) \big(  \coth \frac{y}{\k}-\k\, \coth y \big) 
+     16 e^{- { t}  } \big(\frac{1}{ \sinh \frac{y}{\k}}-\frac{\k}{\sinh y}\big) \Big] \ . \ \ \ \ 
\eea
Here   we isolated the  contribution $ \k\,  \tG_1(1)$ 
 of the first ($AdS_2$) terms in \rf{28} and \rf{42}  where   $  \tG_1(1)$  given by  \rf{4}
 was already computed in \ci{bt}. 
 The first  term in the bracket  in $H$   is the combined contribution of  5+3  scalars   and the second term  is  the contribution of  8 fermions
 (taken with the minus sign as required in \rf{b1}). 
   Both  vanish for $\k=1$, i.e. $H(t,y;1)= 0$. 
   
   The  proper time  integral    is actually convergent at $t\to 0$  as one can see from the 
   small $t$ expansion of the heat kernels in \rf{31a} and \rf{45}
   or directly from \rf{522},\rf{52}   after  setting $y \to y= \sqrt t\, u$
   to  get 
   $\int _{0}^{\infty} {dy}  \ { e^{-\frac{y^{2}}{t}} \ov (4\pi t)^{{1}/{2}}  } \to \int _{0}^{\infty} {du}  \ { e^{- u^2} \ov (4\pi)^{{1}/{2}}  }$:
   \be 
   H(t, \sqrt t \, u; \k)\Big|_{t\to 0} = - {\te {2\ov 15}}( \k  - \k^{-3})  u^2  \, t    +   O(t^2)   \ . \la{53} 
   \ee
The resulting  cancellation of 2d UV  divergences in  \rf{522} is an important check of the consistency 
of this result as the (non-trivial part of the)  1-loop  correction in  \adss    superstring   should be finite on generic 
2d background \ci{dgt}. 

We can thus remove the  proper-time  (UV)  cutoff $\varepsilon$ in \rf{522}
  and compute  the finite  integral over $t$  
(keeping $t$ and $y$  as integration variables and  using  that 
$ \int _{0}^{\infty}\frac{dt}{t^{3/2}} e^{-( \mu^2 t+\frac{y^{2}}{t})}=
 \frac{\sqrt{\pi}}{y} e^{-2\mu y}$)
 \be 
 \la{54}
 \II(\k) = -  \frac{1}{4} \int_{0}^{\infty} {
  dy\ov y \, \sinh y} \Big[ (  5 e^{-y} + 3 e^{-3y} ) \big( \coth \frac{y}{\k}-\k\coth y \big) 
  +  16   e^{-2y}  \big( { 1 \ov \sinh { y \ov \k} }  - { \k  \ov \sinh y} \big) \Big] \ . 
  \ee
  This  integral is  convergent at  both $y\to \infty$ and $y\to 0$   as for small $y$ 
   the integrand  goes as $ {1\ov 6} ( \k - \k^{-1})   + O(y)$. 
   
    It is not clear how to compute the integral \rf{54}  analytically.  For some   integer  values of $\k$ it  is  given by 
 a combination of values of  the Riemann $\zeta$-function and polygamma functions, but we could not find 
  an   interpolating  expression. It is of course  straightforward to evaluate it numerically  
  and conclude that for $\k=k\not=1$ 
   it is different from 
 the earlier result  \rf{b2}  of \ci{kt}     and also from  the  gauge-theory $ { 3\ov 2} \ln k$     behaviour in \rf{b3}. 
 We illustrate  the difference   between $\tG_{1} (k)$ in \rf{51}  and   $\tG_{1, \rm KT} (k)$   in  \rf{b2} 
 in Figure 1. 
 For large $\k$   we  find that  $\tG_{1} $ grows linearly,   $\tG_{1}(\k\gg 1 )\to   1.077 \k $, while for small 
$\k$ we get   $\tG_{1}(\k\to 0 )  \to - 0.15 \k^{-1} $. 
 \begin{figure}[!h]
\centering
\includegraphics[scale=0.6, angle=0]{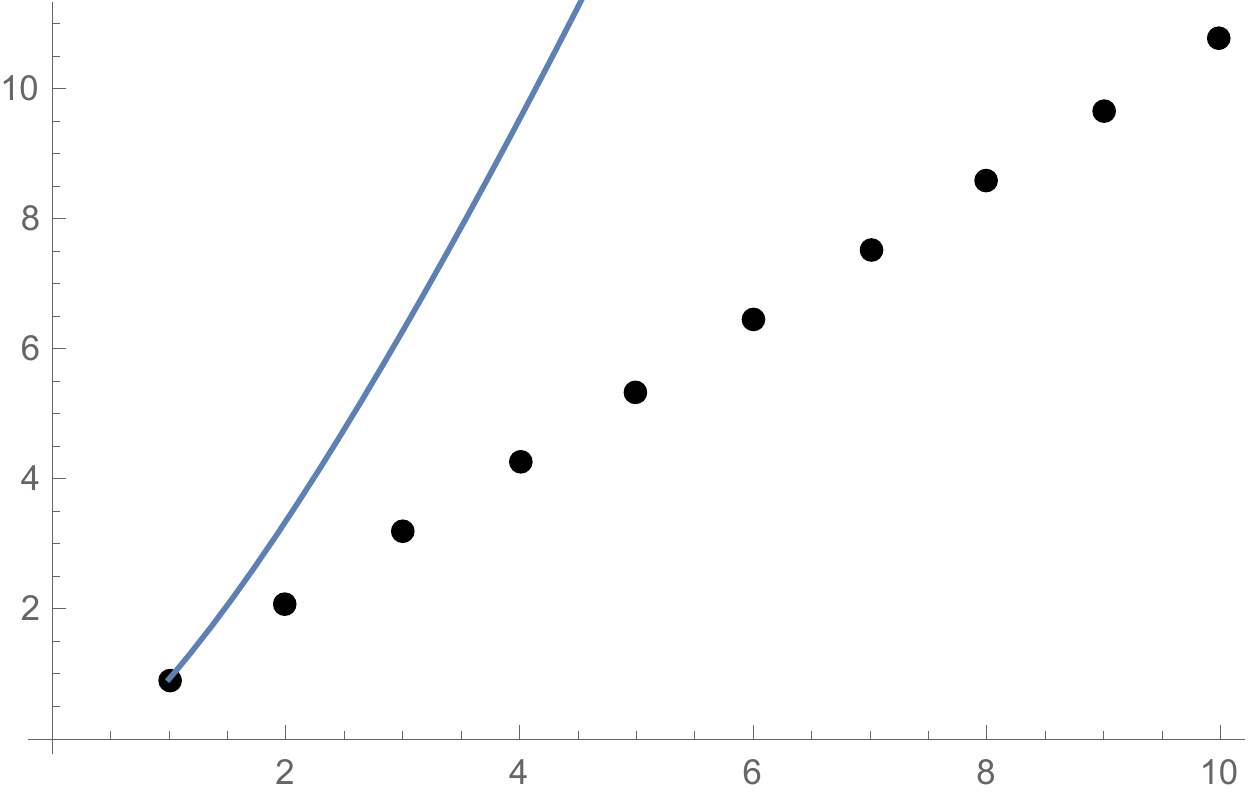}
\caption{Comparison between $\tG_{1}(k)$ in \rf{51}  (black dots)   and   $\tG_{1\,  \rm KT} (k)$   in  \rf{b2}  
(blue line)}
\end{figure}


\section*{Acknowledgments}
A.A.T.  thanks   E. Buchbinder for  a  collaboration at  an initial stage  and useful discussions. 
He  also  acknowledges discussions with 
 V. Forini, D. Fursaev,  M. Kruczenski,   S. Solodukhin  and A. Tirziu.   
The  work of A.A.T. is supported by the ERC Advanced grant No.290456,
 the STFC  grant ST/J0003533/1
and  by  the  Russian Science Foundation grant 14-42-00047  associated with Lebedev Institute.

\appendix
\section{Heat kernels of Laplacians   on   2d  cone  }
\def\theequation{A.\arabic{equation}}
\setcounter{equation}{0}

 Let us consider a $2$-dimensional space   with  a radial coordinate $\xi$ and
  a polar angle $\phi$ of period $\beta =2\pi\k $ with  metric near $\xi=0$   being 
  $ds^2 = d \xi^2 + \xi^2 d \phi^2$ so that  there is a conical singularity with angular deficit $\delta= 2\pi-\beta= 2\pi (1-\k)$.
  Given  the expression for the   heat kernel of a scalar or spinor Laplacian on  a regular manifold ($\k=1$)  it is 
  possible   to derive its counterpart   on the  space with $\k\not=1$    using so called  Sommerfeld formula 
  \ci{dowk,furs}.  We shall review this construction   below. 
  

\subsection{Scalar  case }
Let us  consider first  the case of flat  2d cone 
\begin{equation}\la{a1} 
ds^{2}= d\xi^{2} + \xi^{2}d\phi^{2} \ ,\ \ \ \ \ \ \ \ \ \ \ \phi +2\pi \k  \equiv  \phi \   . 
\end{equation}
To construct the  heat kernel  for $\Delta =- \del^\m\del_\m$  on this cone   it is sufficient to consider  the 
 two points $x(\phi )= (\xi,\phi)$ and $x(0)=(\xi,0)$ separated  only in the  angular   direction  by angle $\phi$.
 The  starting point   is the heat kernel for $\k=1$, i.e. on a  regular $R^2$ plane, where 
    we have  for $K(x,x'; t) = \bra  x| e^{- t\Delta } | x'   \ket $
\begin{equation}\la{a2} 
K(x(\phi),x(0);t)\equiv K_0 ( \xi, \phi,\  \xi,0;t ) = \frac{1}{4\pi t}\,   e^{ -  { 1 \ov 4t} { |x(\phi) - x(0) |^2 } } =  \frac{1}{4\pi t} \, e^{-{1 \ov 2t}(1-\cos \phi){ \xi^2 }}\ . 
\end{equation}
 Since this  is a smooth function of  $\phi$, we can rewrite it using the Cauchy representation:
\begin{equation}\la{a22} 
K(x(\phi),x(0);t)= \frac{1}{2\pi i}\oint_{C}  \frac{dz}{z-\phi}K(x(z),x(0);t)\ ,
\end{equation} 
where $C$ is a contour in the complex plane which encircles the point $z=\phi$.
Now  let us  deform $C$  to a family of contours 
$ A_{n},\ n=0, \pm1,\pm2,....$,   where  $A_{n}$ goes from $ (2n+1)\pi-\epsilon + i\infty$ to $ (2n-1)\pi + \epsilon + i \infty $ in the higher half plane and from $ (2n-1)\pi + \epsilon - i\infty$ to $ (2n+1)\pi - \epsilon - i \infty $ in the lower 
half plane 
one. Then  the heat kernel can be expressed as the ``sum over images"
\bea
 K(x(\phi),x(0);t)  = \frac{1}{2\pi i}\sum_{n=-\infty}^{\infty} \int_{A_{n}}\frac{dz}{z-\phi}K(x(z),x(0);t)  \no \qquad \qquad \\
 = \frac{1}{2\pi i} \sum_{n=-\infty}^{\infty} \int_{A_{0}}\frac{dz}{z-\phi -2\pi n}K(x(z),x(0);t) 
 =\frac{1}{4\pi i} \int_{A_{0}} dz\,  \cot \frac{z-\phi}{2} \  K(x(z),x(0);t)\ .  \la{a3}
 \eea 
 Here we used the periodicity of the kernel in \rf{a2}   and 
the identity $
\sum _{n=-\infty}^{\infty} \frac{1}{z-n q}=\frac{\pi}{q} \cot \frac{\pi z}{q}\ .$

We can  now repeat the same procedure changing the period
 of the angle from $2\pi $ to $2 \pi\k$   to get the heat  kernel on the cone \rf{a1} 
 \begin{equation}\la{a4} 
K_{\k}(x(\phi),x(0);t)  =\frac{1}{4\pi i \k } \int_{A_{0}} dz\,  \cot \frac{z-\phi}{2\k} \  K (x(\phi),x(0);t) \    , 
\end{equation}
or  after  shifting  $z\rightarrow z +\phi$
\begin{equation}\la{a5} 
K_{\k }(x(\phi),x(0);t)= \frac{1}{4\pi i  \k } \int_{A_{0}^{\prime}} dz\,  \cot  \frac{ z}{2\k }\,   K(x(z+\phi),x(0);t)\ . 
\end{equation}
The contour $A_{0}^{\prime}$ is obtained  from  $A_{0}$ by a 
 shift by $\phi$ and consists of two pieces: one in the upper half plane running from $\pi + \phi+ i\infty$ to $-\pi +\phi +i\infty$ and one in the lower half plane going from $-\pi +\phi -i\infty$ to $\pi+\phi-i\infty$. 
We may  then  deform $A_{0}^{\prime}$ into  a small circle around $z=0$
 and another contour $\Gamma^{\prime}$  made of two vertical lines: 
 one going from $ -\pi + \phi -i \infty $ to $-\pi +\phi+ i \infty$ and one from $ \pi +\phi +i \infty $ to $\pi +\phi -i \infty$.
  Near $z=0$ the integrand has a pole of $\cot\frac{ z}{2 \k }\sim \frac{2 \k }{ z}$ with residue
  being the  regular plane kernel $ K(x(z+\phi),x(0);t)$. 
  Then  denoting $ \Gamma= -\mathrm{\Gamma^{\prime}}$ we finish   with 
\begin{equation}
\la{a6}
 K_{\k}(x(\phi),x(0);t)=K(x(\phi),x(0);t) + \frac{i}{4\pi \k} \int_{\Gamma} dz\,  \cot  \frac{z}{2 \k}\ K(x(z+\phi),x(0);t) \ . 
\end{equation}
This is the Sommerfeld formula for the  heat kernel of the 
scalar Laplacian  on the cone \rf{a1}  in terms of the  heat kernel on the  regular 2-plane. 

 The same procedure can be applied to the case of an arbitrary  space  $\mathcal{M}$ where the heat kernel depends only on 
 the angular distance between two points. Denoting  $x=(\xi,\phi)$, $x'=(\xi^{\prime},\phi^{\prime})$ the coordinates of two 
  points on the  corresponding cone  $\mathcal{M}_{\k }$  of $\mathcal{M}$, 
  the Sommerfeld formula for the heat kernel of the scalar Laplacian reads 
\begin{equation}
\la{a7}
 K_{\k}(\xi,\xi^{\prime},\phi,\phi^{\prime};t)= K(\xi,\xi^{\prime},\phi,\phi^{\prime};t) 
  + \frac{i}{4\pi \k} \int_{\Gamma} dz\,  \cot  \frac{z}{2 \k}\  
 K(\xi,\xi^{\prime},\phi +z,\phi^{\prime} ;t).
\end{equation}
Here the angular distance is  $z=\phi-\phi^{\prime}$ and the contour $\Gamma$ 
 is formed by  two vertical lines 
  from $ -\pi + (\phi-\phi^{\prime}) +i \infty $ to $-\pi +(\phi-\phi^{\prime})- i \infty$ and from $ \pi +(\phi-\phi^{\prime}) -i \infty $ to $\pi +(\phi-\phi^{\prime}) +i \infty$.
Taking the coincident points   limit  we get 
\begin{equation}
\la{a8}
 K_{\k}(\xi=\xi^{\prime},\phi=\phi^{\prime};t)=
  K(\xi=\xi^{\prime},\phi=\phi^{\prime};t)+\frac{i}{4\pi \k} \int_{\Gamma} dz\,  \cot  \frac{z}{2 \k}
 \ K(\xi=\xi^{\prime},\phi^{\prime}=\phi+z;t)\ . 
 \end{equation}
Defining   the integrated heat kernel
\begin{equation}\la{a88}
K(z;t)= \int_{M_{\k }}d\phi \, d\xi \, \sqrt{g} \ K(\xi=\xi^{\prime},\phi^{\prime}=\phi+z;t)\ ,
\end{equation}
we finally get the  expression for the trace  of  heat kernel  
\bea \la{a9}
 \KK_\k (t)=\int_{M_\k} d^2 x\,  \sqrt g \, \bra  x| e^{- t\Delta } | x  \ket =
  K(0; t) + \frac{i}{4\pi \k} \int_{\Gamma} dz\,  \cot  \frac{z}{2\k}\ K(z;t)\  . 
\eea 
Since in \rf{a9}  $\phi=\phi'$, here  the contour $\Gamma$ reduces to
 two vertical lines going from $ -\pi -i \infty $ to $-\pi + i \infty$ and from $ \pi - i\infty $ to $\pi +i \infty$.
 
Thus in addition to the expected term $K(0; t)$   ``inherited" from the  regular space case 
there is  a correction  given  by a 
 contour integral on the complex plane.
 For $\k=1$  this second term vanishes as expected: 
  the integrand becomes $2\pi$ periodic in $z$ 
   and the contributions coming from the integration along the two vertical lines at  $-\pi$ and $\pi$ cancel each other. 

\subsection{Spinor   case}

Let us start again with a   flat space cone.  
On $R^2$   in Cartesian coordinates   the spin connection is trivial and the fermionic Laplace operator 
$ \Delta_{\hal}= -{\nabla}^{2}+\frac{1}{4}{R} $ reduces to the scalar Laplacian $\Delta_0 = -\partial^{\mu}\partial_{\mu}$. 
Thus  solution to the heat kernel equation is the same as for the scalar  \rf{a2}  
up to $2\times 2$ identity matrix $I$:
\begin{equation}\la{a10}
  \td K\left(x,x';t\right)= K_{0}(x,x';t)\,  I \ . 
\end{equation}
In polar  coordinates  $( \xi , \phi )$
 the spinor  covariant derivative is   $\nabla_{\mu} = \partial_{\mu} + \omega_{\mu}$
  with $\omega_{\mu}dx^{\mu}= -\frac{i}{2}\sigma_{3}d\phi$. 
  Thus  
 \begin{equation} \la{a11} 
  \nabla_{\mu} =  \V \partial_{\mu} V^{-1} \ , \ \ \ \ \ \ \ \ \ 
  \Delta_\hal =   \V \Delta_0   \V^{-1} \ , \qquad \qquad  \V=e^{ \frac{i}{2}\sigma_{3}\phi} \ . 
\end{equation}
Then 
the solution to the spinor  heat kernel equation in polar coordinates  may be written as 
\begin{equation}\la{a12} 
 K(x,x';t) = \V(\phi)\, \td K(x,x';t) \ , 
\end{equation}
with $\td K$  given by \rf{a10}. 
Due  to the presence of the  spin factor $\V$  the spinor   heat kernel  is antiperiodic  in $\phi$.

To get spinor heat kernel on a cone of $R^2$  we   follow the same  route  as in  the scalar case:
 we choose $x(\phi) =(\xi,\phi)$, $x(0) =(\xi,0)$  and  first  rewrite $ K(x,x';t)$  in \rf{a12} 
  using 
  the method of the images   as in \rf{a22},\rf{a3}: 
\bea \no   \la{a13} 
&&K (x(\phi),x(0);t)= \frac{1}{2\pi i}\oint_C  { \frac{dz}{z-\phi}\,  \V(z)\, K(x(z),x(0);t)} \no \\
&&\qquad\qquad\qquad \quad   =  \frac{1}{2\pi i}\sum_{n=-\infty}^{\infty} \int_{A_{n}}\frac{dz}{z-\phi}\ \V(z)\,  K(x(z),x(0);t) \no \\
 &&\qquad\qquad\qquad \quad= \frac{1}{2\pi i} \sum_{n=-\infty}^{\infty} \int_{A_{0}}\frac{dz\, (-1)^{n}}{z-\phi -2\pi n}\, \V(z)\, K(x(z),x(0);t)\no \\
 &&\qquad\qquad\qquad \quad=\frac{1}{4\pi i} \int_{A_{0}} \, dz \,  \frac{1}{\sin \frac{z-\phi}{2} }\, \V(z) \, K(x(z),x(0);t) \ . \la{a15}
\eea 
 Here we used  the antiperiodicity  of $\V$   and 
$
\sum _{n=-\infty}^{\infty} \frac{(-1)^{n}}{z-q n }=\frac{\pi}{q} \frac{1}{\sin\frac{\pi z}{q}}$. 
We can now  change the periodicity of  $\phi$ from $2\pi $ to $2\pi \k   $ and 
to get   the  expression for the spinor heat kernel on the cone of $R^2$
\bea
K_{ \k }(x(\phi),x(0);t)&=& \frac{1}{4\pi i \k } \int_{A_{0}} \,dz\ \frac{1}{\sin\frac{z-\phi}{2 \k }}\, K(x(z),x(0);t)\no\\
 &=&
\frac{1}{4\pi i \k } \int_{A_{0}^{\prime}}\, dz \  \frac{1}{\sin\frac{z}{2\k }}\, K(x(z+\phi),x(0);t)\ , \la{a14}
\eea
where $A_{0}^{\prime}$ is again given by two  lines  in the upper and lower half plane:  one running from $\pi + \phi+ i\infty$ to $-\pi +\phi +i\infty$ and the  other   from $-\pi +\phi -i\infty$ to $\pi+\phi-i\infty$.
Next,   we deform $A_{0}^{\prime}$ into  a small circle around $z=0$
 and a contour $\Gamma^{\prime}$ made of two vertical lines that run from $ -\pi + \phi -i \infty $ to $-\pi +\phi+ i \infty$ and from $ \pi +\phi +i \infty $ to $\pi +\phi -i \infty$.  Setting  $ \Gamma= -\Gamma^{\prime}$ we  obtain
\begin{equation}\la{a16}
 K_{\k}(x(\phi),x(0);t)= K(x(\phi),x(0);t)+
  \frac{i}{4\pi \k} \int_\Gamma\, dz \  \frac{1}{\sin\frac{z}{2 \k}   }\,  K(x(z+\phi),x(0);t) \   . 
\end{equation}
This expression can be generalized to the case of a  cone of    curved 2d  space 
 where the heat kernel depends on the angular distance between two points
 $x=(\xi,\phi)$ and $x'=(\xi^{\prime},\phi^{\prime})$ 
 \begin{equation}
K_{\k}(\xi,\xi^{\prime},\phi,\phi^{\prime};t)=K(\xi,\xi^{\prime},\phi,\phi^{\prime};t) + \frac{i}{4\pi \k} \int_{\Gamma}  \, dz
 \frac{1}{\sin  \frac{z}{2\k}  }\,  K(\xi,\xi^{\prime},\phi+z,\phi^{\prime};t)\ . \la{a17} 
\end{equation}
Here, as in the scalar  case in \rf{a7}, 
 $\Gamma$  is modified   by  $\phi \rightarrow \phi-\phi^{\prime}$.
Defining  the integrated  kernel as in \rf{a88}   we then get
 the final expression for the traced spinor heat kernel:
\begin{equation}\la{a20} 
\KK_{\k}(t)= \mathrm{Tr}\, K (0;t) + \frac{i}{4\pi \k} \int_{\Gamma}\,  dz \  \frac{1}{\sin\frac{z}{2\k}} \, \mathrm{Tr}\, K(z;t)\  . 
\end{equation}
Here $\Tr$ is  over the spinor indices and the integration contour reduces for $\phi=\phi^{\prime}$ to the  two vertical lines 
going from $ -\pi -i \infty $ to $-\pi + i \infty$ and from $ \pi - i\infty $ to $\pi +i \infty$.
For  $\k=1$ the integrand becomes  periodic (as the product of two antiperiodic functions in $z$)
and thus  the integral contribution vanishes.

\section{Matrix  $U$  in   spinor heat kernel on  $AdS_2$ }
\def\theequation{B.\arabic{equation}}
\setcounter{equation}{0}
Here   we   solve the equation \rf{32}  for  the matrix $U$  in 
the $AdS_2$ metric  \rf{aa}  
written in  polar coordinates.
The     corresponding zweibein  and  tangent  space    and spinor  connections   are 
 ($\mu=( \xi,\phi )$)
\bea
&&  e^{1}_{\mu}=(1,0)\ , \ \ \ \ \ e^{2}_{\mu}=(0,\sinh \xi)\ , \ \ \ \ \ \  \ \  \omega_{\xi}^{\a \beta}=0\ , \ \   
\ \ \ \    \omega_{\phi}^{12}=-\omega_{\phi}^{21}=-\cosh \xi \ , \la{33a}\\
&& \omega_\m  = { 1  \ov 2}  \omega_{\mu}^{\a \beta}  \Sigma_{\a\b}
 =(0,-\frac{i}{2}\cosh \xi\ \sigma_{3})\ , \ \ \ \  \ \ \ \    \Sigma_{\a\b}  = {1\ov 4} [ \g_\a , \g_\b  ] \ , \ \ \ \ \ \ 
\gamma^{\a}=( \sigma_{1},\s_2) \ . \la{3322} 
\eea
In \cite{camp} it was  shown that  (\ref{32}) is equivalent to 
the requirement  that the spinor covariant derivative  acts on $U(\xi,\xi^{\prime},\phi,\phi^{\prime})$ in the following way
\begin{equation}
\la{34b}\te 
 \nabla_{\m}U=  \partial_{\m}U +\omega_{\m}\, U=
  A\,  \Sigma_{\a \beta}\, n^{\beta}\,  U\ , \ \ \ \ \ \ \ \
A\equiv \tanh\frac{d(\xi,\xi^{\prime},\phi,\phi^{\prime})}{2}\ , 
\end{equation} 
where   the  geodesic distance between the two points $(\xi,\phi)\ , (\xi^{\prime},\phi^{\prime})$ is 
given in \rf{aaa}.
This  equation for  $U$  can be rewritten  as
\be\la{35b}
 \partial_{\xi}U=\frac{i}{2}{ \tanh {d\ov 2 } \ov {\sinh\xi }\,  } {\partial_{\phi}d}\, \sigma_{3}U\ , \qquad 
{ \partial_{\phi}U}=\frac{i}{2}{\sinh\xi} \big(  \coth\xi-\tanh \frac{d}{2}\,  \partial_{\xi}d\big) \,   \sigma_{3} U \ . 
\ee
The first equation in \rf{35b} 
has solution 
\bea\label{36b}
&& 
U(\xi,\xi^{\prime},\phi,\phi^{\prime})=
 \exp \big[ \frac{i}{2}\sigma_{3}B(\xi,\xi^{\prime},\phi,\phi^{\prime})\big] \,  C(\phi,\phi^{\prime}) \ ,\\
&&
B(\xi,\xi^{\prime},\phi,\phi^{\prime})= \int {d\xi \ov \sinh \xi}\ \partial_{ \phi} \log \big[
1 + \cosh d(\xi,\xi^{\prime},\phi,\phi^{\prime})\big] \ . \la{37a}
\eea 
 Using \rf{aaa}  we find
\bea
B&=& \int { d\xi \ov \sinh\xi}\ \frac{\sinh\xi\sinh\xi^{\prime}\sin(\phi-\phi^{\prime})}{1+\cosh\xi\cosh\xi^{\prime}-\sinh\xi\sinh\xi^{\prime}\cos(\phi-\phi^{\prime})} \no 
\\ &=& 
2  \arctan \frac{\cosh \frac{\xi+\xi^{\prime}}{2}\ \tan\frac{\phi-\phi^{\prime}}{2}}{\cosh \frac{\xi-\xi^{\prime}}{2}} -{\pi} \ .  \la{3888}
\eea 
The   matrix
$C(\phi,\phi^{\prime})$  can be determined  from  
 the second equation
in \rf{35b}
\be\la{39b}
\partial_{\phi}C=
 \frac{i}{2}\sigma_{3} \big( \cosh\xi-\sinh \xi \tanh \frac{d}{2} \,  \partial_{\xi} d-\partial_{\phi}B \big) C =0 \ , 
\ee
where we used  \rf{aaa}  and \rf{3888}. Thus $C$ is constant and  is fixed by 
 the initial condition $U(\xi=\xi^{\prime},\phi=\phi^{\prime})=I$
 \begin{equation}\la{40b}
C(\phi,\phi^{\prime})=  e^{  
\frac{i}{2}\pi \sigma_{3}   }  = i \s_3   \ . 
\end{equation}
As a result, 
\begin{equation}
U(\xi,\xi^{\prime},\phi,\phi^{\prime})= \exp\Big(  i \sigma_{3}\,  \arctan \frac{\cosh \frac{\xi+\xi^{\prime}}{2}\, \tan\frac{\phi-\phi^{\prime}}{2}}{\cosh \frac{\xi-\xi^{\prime}}{2}} \Big)\  . \la{uub}
\end{equation}

\



\begin{thebibliography}{99}
\baselineskip 14pt
\bibitem{bt} 
  E.~I.~Buchbinder and A.~A.~Tseytlin,
  ``1/N correction in the D3-brane description of a circular Wilson loop at strong coupling,''
  Phys.\ Rev.\ D {\bf 89}, no. 12, 126008 (2014)
  [arXiv:1404.4952].


\bibitem{ber} 
  D.~E.~Berenstein, R.~Corrado, W.~Fischler and J.~M.~Maldacena,
  ``The Operator product expansion for Wilson loops and surfaces in the large N limit,''
  Phys.\ Rev.\ D {\bf 59}, 105023 (1999)
  [hep-th/9809188].


\bibitem{dgo} 
  N.~Drukker, D.~J.~Gross and H.~Ooguri,
  ``Wilson loops and minimal surfaces,''
  Phys.\ Rev.\ D {\bf 60}, 125006 (1999)
  [hep-th/9904191].
  
  
\bibitem{Erickson:2000af} 
  J.~K.~Erickson, G.~W.~Semenoff and K.~Zarembo,
  ``Wilson loops in N=4 supersymmetric Yang-Mills theory,''
  Nucl.\ Phys.\ B {\bf 582}, 155 (2000)
  [hep-th/0003055].
G.~W.~Semenoff and K.~Zarembo,
  ``Wilson loops in SYM theory: From weak to strong coupling,''
  Nucl.\ Phys.\ Proc.\ Suppl.\  {\bf 108}, 106 (2002)
  [hep-th/0202156].


\bibitem{Drukker:2000rr}
  N.~Drukker and D.~J.~Gross,
  ``An Exact prediction of N=4 SUSYM theory for string theory,''
  J.\ Math.\ Phys.\  {\bf 42} (2001) 2896
  [hep-th/0010274].
  


\bibitem{Drukker:2005kx}
  N.~Drukker and B.~Fiol,
  ``All-genus calculation of Wilson loops using D-branes,''
  JHEP {\bf 0502} (2005) 010
  [hep-th/0501109].

\bibitem{Pestun:2007rz} 
  V.~Pestun,
  ``Localization of gauge theory on a four-sphere and supersymmetric Wilson loops,''
  Commun.\ Math.\ Phys.\  {\bf 313}, 71 (2012)
  [arXiv:0712.2824].
  




\bibitem{dgt}
  N.~Drukker, D.J.~Gross and A.A.~Tseytlin,
  ``Green-Schwarz string in AdS(5) x S5: Semiclassical partition function,''
  JHEP {\bf 0004} (2000) 021
  [hep-th/0001204].
 
\bi{thei}
  S.~Forste, D.~Ghoshal and S.~Theisen,
  ``Stringy corrections to the Wilson loop in N=4 superYang-Mills theory,''
  JHEP {\bf 9908}, 013 (1999)
  [hep-th/9903042].

 \bi{cha}
 I. Chavel, ``Heat kernels and spectral theory",  Cambridge U.P., 1984. 

\bibitem{cam}
  R.~Camporesi,
  ``Harmonic analysis and propagators on homogeneous spaces,''
  Phys.\ Rept.\  {\bf 196} (1990) 1.

\bibitem{Fursaev:2011zz} 
  D.~Fursaev and D.~Vassilevich,
  ``Operators, Geometry and Quanta : Methods of spectral geometry in quantum field theory,''
  D.~V.~Vassilevich,
  ``Heat kernel expansion: User's manual,''
  Phys.\ Rept.\  {\bf 388}, 279 (2003)
  [hep-th/0306138].


  
\bibitem{kt}
  M.~Kruczenski and A.~Tirziu,
  ``Matching the circular Wilson loop with dual open string solution at 1-loop in strong coupling,''
  JHEP {\bf 0805} (2008) 064
  [arXiv:0803.0315].


\bibitem{km}
  C.~Kristjansen and Y.~Makeenko,
  ``More about One-Loop Effective Action of Open Superstring in $AdS_5\times S^5$,''
  JHEP {\bf 1209} (2012) 053
  [arXiv:1206.5660].
  
  \bi{sol}
  S.~N.~Solodukhin,
  ``Entanglement entropy, conformal invariance and extrinsic geometry,''
  Phys.\ Lett.\ B {\bf 665}, 305 (2008)
  [arXiv:0802.3117].
  H.~Casini and M.~Huerta,
  ``Entanglement entropy for the n-sphere,''
  Phys.\ Lett.\ B {\bf 694}, 167 (2010)
  [arXiv:1007.1813 [hep-th]].
  I.~R.~Klebanov, S.~S.~Pufu, S.~Sachdev and B.~R.~Safdi,
  ``Renyi Entropies for Free Field Theories,''
  JHEP {\bf 1204}, 074 (2012)
  [arXiv:1111.6290].
  R.~Aros, F.~Bugini and D.~E.~Diaz,
  ``On Renyi entropy for free conformal fields: holographic and q-analog recipes,''
  J.\ Phys.\ A {\bf 48}, 105401 (2015)
  [arXiv:1408.1931].
  
 
\bibitem{ms} 
  R.~B.~Mann and S.~N.~Solodukhin,
  ``Universality of quantum entropy for extreme black holes,''
  Nucl.\ Phys.\ B {\bf 523}, 293 (1998)
  [hep-th/9709064].


  \bi{jon}  T.H. Jones and  D. Kucerovsky, 
  ``Heat Kernel for Simply-Connected Riemann Surfaces'', 
  arXiv:1007.5467 [math.DG]. \\
  T.H. Jones, ``The heat kernel on noncompact Riemann surfaces", 
  PhD thesis (2008). 
 
 
   \bi{gul}
  R.~K.~Gupta, S.~Lal and S.~Thakur,
 ``Heat Kernels on the $AdS_2$ cone and Logarithmic Corrections to Extremal Black Hole Entropy,''
  JHEP {\bf 1403} (2014) 043
  [arXiv:1311.6286].
   ``Logarithmic Corrections to Extremal Black Hole Entropy in N = 2, 4 and 8 Supergravity,''
  arXiv:1402.2441.

 
  
\bi{dowk}
J.~S.~Dowker,
  ``Quantum Field Theory on a Cone,''
  J.\ Phys.\ A {\bf 10}, 115 (1977).
  ``Vacuum Averages for Arbitrary Spin Around a Cosmic String,''
  Phys.\ Rev.\ D {\bf 36}, 3742 (1987).
  D.~V.~Fursaev,
  ``The Heat kernel expansion on a cone and quantum fields near cosmic strings,''
  Class.\ Quant.\ Grav.\  {\bf 11}, 1431 (1994)
  [hep-th/9309050].
  J.~S.~Dowker,
  ``Effective actions with fixed points,''
  Phys.\ Rev.\ D {\bf 50}, 6369 (1994)
  [hep-th/9406144].
  
  
  \bi{camp}  R.~Camporesi,  ``The Spinor heat kernel in maximally symmetric spaces,''
  Commun.\ Math.\ Phys.\  {\bf 148} (1992) 283.
  R.~Camporesi and A.~Higuchi,
  ``Spectral functions and zeta functions in hyperbolic spaces,''
  J.\ Math.\ Phys.\  {\bf 35} (1994) 4217.
  

\bi{camf}
 R.~Camporesi and A.~Higuchi,
  ``On the Eigen functions of the Dirac operator on spheres and real hyperbolic spaces,''
  J.\ Geom.\ Phys.\  {\bf 20}, 1 (1996)
  [gr-qc/9505009].


\bi{kab}
  D.~N.~Kabat,
  ``Black hole entropy and entropy of entanglement,''
  Nucl.\ Phys.\ B {\bf 453}, 281 (1995)
  [hep-th/9503016].
  
  

\bi{fm}
D.~V.~Fursaev and G.~Miele,
  ``Cones, spins and heat kernels,''
  Nucl.\ Phys.\ B {\bf 484}, 697 (1997)
  [hep-th/9605153].

\bibitem{furs} 
  D.~V.~Fursaev,
  ``Euclidean and canonical formulations of statistical mechanics in the presence of killing horizons,''
  Nucl.\ Phys.\ B {\bf 524}, 447 (1998)
  [hep-th/9709213].



 \end{thebibliography}
\end{document}